\documentclass[aps,prb,twocolumn,groupedaddress,amsmath,amssymb,10pt]{revtex4}
\usepackage{graphicx,bm,wasysym,color}

\def\be{\begin{equation}}  
\def\ee{\end{equation}}  
\def\ba{\begin{eqnarray}}  
\def\ea{\end{eqnarray}}  
\def\bc{\begin{center}}  
\def\ec{\end{center}}

\begin{document}
%\large
\title{Comment on \textit{``Graphene---A rather ordinary nonlinear optical material''} [Appl. Phys. Lett. \textbf{104}, 161116 (2014)]}

\author{S. A. Mikhailov}
\email[Email: ]{sergey.mikhailov@physik.uni-augsburg.de}
\affiliation{Institute of Physics, University of Augsburg, D-86135 Augsburg, Germany}

\date{\today}

%\begin{abstract}
%\end{abstract}

%\pacs{78.67.Wj, 78.45.+h, 41.60.-m}

\maketitle

In 2007 it was predicted  \cite{Mikhailov07e} that, due to the linear energy dispersion of quasi-particles in graphene, this material should have a strongly nonlinear electrodynamic response. This prediction stimulated a number of theoretical and experimental studies in which harmonics generation, four-wave mixing and other nonlinear phenomena have been observed. However, in April 2014 Khurgin published a paper \cite{Khurgin14} entitled ``Graphene---A rather ordinary nonlinear optical material''. In that paper the author performed a general qualitative analysis of the nonlinear properties of graphene, made order-of-magnitude estimates of nonlinear graphene parameters and came to a conclusion that practical nonlinear optical devices based on graphene will have no particular advantage over other materials. Therefore, a question arises whether it makes sense to continue research on the nonlinear graphene physics, or the analysis in Ref. \cite{Khurgin14} was not completely correct. 

In April 2014 the theory of the nonlinear electrodynamic response of graphene was not sufficiently developed, therefore the Khurgin's arguments and estimates, which looked quite reasonable, could not be confirmed or refuted. Nowadays more theoretical results are available, including a comprehensive quantum theory of the third-order nonlinear electrodynamic response of graphene \cite{Cheng15,Mikhailov16a}, so it is possible to verify the validity of the Khurgin's conclusions.

In the beginning of Ref. \cite{Khurgin14} the author discussed the high-frequency (optical) response of graphene, determined by the \textit{inter-band} effects. Mainly talking about all optical switching (AOS) in his paper, for which the parameter 
\be 
\frac{\sigma^{(3)}E^2}{\textrm{Re}~\sigma^{(1)}}
\equiv \zeta
\frac{E^2}{E_i^2},
\label{ratio31}
\ee
should be about $\pi$, he estimated the factor $\zeta$ as
\be 
\zeta\equiv \frac{\sigma^{(3)}E_i^2}{\textrm{Re}~\sigma^{(1)}}\approx \frac 13\ \ \ \textrm{ in Ref. }\cite{Khurgin14};
\label{estimateKhurgin}
\ee
here $\sigma^{(1)}$ and $\sigma^{(3)}$ are the first- and third-order conductivities, $E_i=\hbar\omega^2/ev_F$, $\omega$ and $E$ are the frequency and the electric field of the wave, and $v_F\approx 10^8$ cm/s is the Fermi velocity in graphene; the estimate (\ref{estimateKhurgin}) was done at the telecommunication wavelength $1.55$ $\mu$m where $E_i\approx 10^7$ V/cm. Further the author writes ``The nonlinear index for the interband transitions can become very large due to the resonant enhancement, but so does the absorption coefficient'' and concludes that graphene is a quite ordinary nonlinear material at optical and near-infrared frequencies since the electric field needed for observation of essential nonlinear effects ($\sim 10^7$ V/cm) is very large.

Let us verify the validity of this conclusion. The third-order nonlinear properties of graphene are characterized by the fourth order tensor $\sigma_{\alpha\beta\gamma\delta}^{(3)} (\omega_1,\omega_2,\omega_3)$ which depends on three input frequencies. At optical wavelengths, when all input frequencies $\hbar\omega_{1,2,3}$ exceed the double Fermi energy $2E_F$ and $\sigma^{(1)}\approx e^2/4\hbar$, the ratio $\sigma^{(3)}/\sigma^{(1)}$ reads \cite{Mikhailov16a,Cheng15}
\begin{widetext}
\be 
\frac{\sigma^{(3)}_{xxxx}(\omega_1,\omega_2,\omega_3)}{\sigma^{(1)}}=\frac {e^2 v_F^2}{ \hbar^2}
\frac{\omega_1+ \omega_2+\omega_3+i3\gamma/2} {(\omega_1+\omega_2+i\gamma)(\omega_2+\omega_3+i\gamma)
(\omega_3+\omega_1+i\gamma)  (\omega_1+\omega_2+\omega_3+i\gamma)^2}.
\ee
\end{widetext}
Here $\gamma$ is the phenomenological relaxation rate. 

Now the parameter (\ref{ratio31}) depends on which physical effect is considered. If we assume that all three input frequencies are the same, $\omega_1=\omega_2=\omega_3=\omega$, which corresponds to the third harmonic generation experiment, we obtain at $\omega\gg \gamma$
\be 
\zeta_+(\omega)=\frac{\sigma^{(3)}_{xxxx}(\omega,\omega,\omega)E_i^2}{\sigma^{(1)}}\approx \frac 1{24}.\label{z+}
\ee
This is even worse (eight times smaller) that the Khurgin's estimate (\ref{estimateKhurgin}). However, for the AOS phenomena a relevant quantity is $\sigma^{(3)}_{xxxx}(\omega,\omega,-\omega)$, with $\omega_1=\omega_2=-\omega_3=\omega$. In this case we get at $\omega\gg \gamma$
\be 
\zeta_-(\omega)=\frac{\sigma^{(3)}_{xxxx}(\omega,\omega,-\omega)E_i^2}{\sigma^{(1)}}=
-\frac {1}{ 2}\frac{\omega^2}{\gamma^2}.\label{z-}
\ee
At optical frequencies $\hbar\omega\simeq 1$ eV while the \textit{inter-band} relaxation rate lies in the meV range \cite{Alexander17}. The required electric field is then about \textit{three orders of magnitude smaller} than the estimate $\sim 10$ MV/cm obtained in \cite{Khurgin14}. 

\begin{figure}
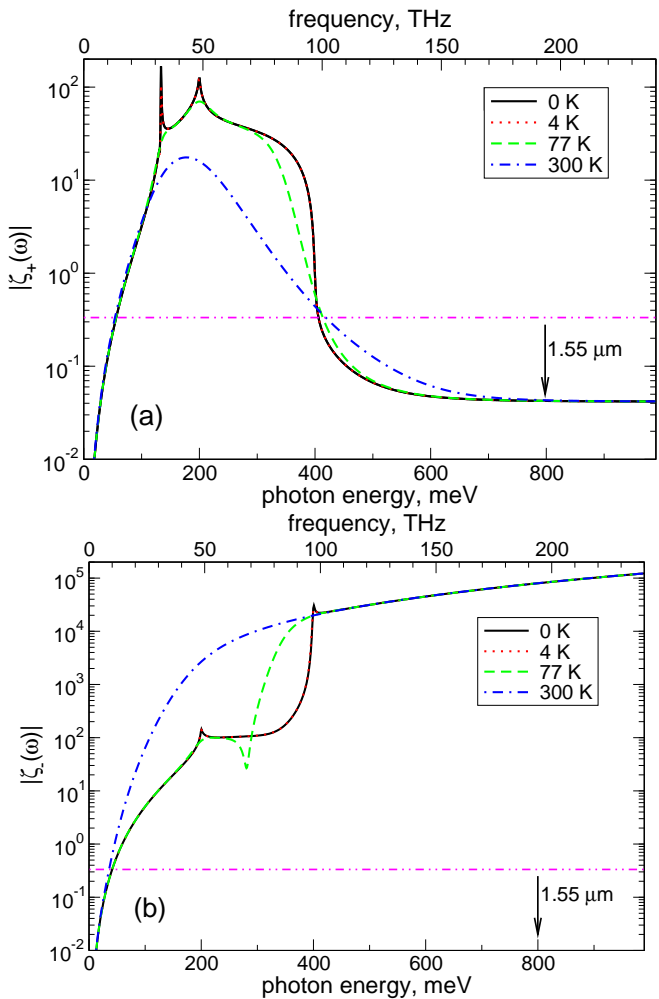

\includegraphics[width=8.65cm]{WWWabsDim.eps}
\includegraphics[width=8.35cm]{WW-WabsDim.eps}
\caption{\label{fig:+-} The frequency dependence of the absolute value of the parameters (a) $\zeta_+(\omega)$, Eq. (\ref{z+}), and (b) $\zeta_-(\omega)$, Eq. (\ref{z-}), at different temperatures. Here $E_F=0.2$ eV (corresponds to the electron density $n_s\approx 3\times 10^{12}$ cm$^{-2}$) and $\hbar\gamma=2$ meV. The magenta dash-double-dotted line shows the estimate (\ref{estimateKhurgin}) from Ref. \cite{Khurgin14}. The black arrow indicates the position of the telecommunication wavelength $\lambda=1.55$ $\mu$m.}
\end{figure}

The statement that the nonlinear parameters of graphene could be substantially increased near inter-band resonances but this is compensated by a similar enhancement of the absorption coefficient is also incorrect. It was shown (see, e.g., a detailed discussion in Ref. \cite{Mikhailov16a}) that the third conductivity $\sigma^{(3)}_{xxxx}(\omega,\omega,\omega)$ responsible for the third harmonic generation has a \textit{second-order pole} at $3\hbar\omega=2E_F$ while the linear conductivity $\sigma^{(1)}(\omega)$ has a step-like behavior in the real part and a \textit{weak logarithmic} resonance in the imaginary part. Figure \ref{fig:+-} illustrates the resonant features of the parameters $\zeta_\pm(\omega)$ in a broader range of frequencies from microwaves up to near-IR. One sees that at frequencies below $\simeq 100$ THz the parameter $|\zeta_+(\omega)|$ exceeds the estimate (\ref{estimateKhurgin}) by almost (more than) two orders of magnitude at room (cryogenic) temperatures, Fig. \ref{fig:+-}(a). The factor $|\zeta_-(\omega)|$ exceeds the estimate (\ref{estimateKhurgin}) at all frequencies above few tens of THz by many orders of magnitude, Fig. \ref{fig:+-}(b).

In the second part of Ref. \cite{Khurgin14} the author discusses the low-frequency regime $\hbar\omega_i\lesssim 2E_F$, where the nonlinear response is determined by the \textit{intra-band} contributions to the conductivity \footnote{The second part of Ref. \cite{Khurgin14} contains some mistakes. The function $F_{\sigma 1}(x)$ cannot well approximate the function $F_{\sigma}(x)$ defined in Eq. (1) of Ref. \cite{Khurgin14} since at large $x\gg 1$ $F_{\sigma}(x)$ behaves as $1/x$ while $F_{\sigma 1}(x)$ as $4/x^2$. As a result, the formula (3) in Ref. \cite{Khurgin14} is incorrect.}. He agrees that at low doping densities the graphene response is highly nonlinear indeed, but remarks that at low densities the overall conductivity is small. The author concludes, in particular, that for operation around 30 THz the nonlinear graphene parameters are though respectable but not superior. 

That graphene nonlinearities should be especially strong at low electron densities was emphasized already in the first publication \cite{Mikhailov07e}. It was shown there that at low frequencies ($\hbar\omega_i\lesssim 2E_F$) the characteristic electric field which determines when the nonlinear effects in graphene become essential ($E\gtrsim E_0$) is given by
\be 
E_0=\frac{\hbar\omega k_F}{e}=\frac{\hbar\omega \sqrt{\pi n_s}}{e},\label{field}
\ee
where $k_F$ is the Fermi wave-vector. As seen from (\ref{field}) the intra-band nonlinearity in graphene will be especially strong at low frequencies and low densities (say, at $f\simeq 1$ THz and $n_s\simeq 10^{11}$ cm$^{-2}$ rather than at $f\simeq 30$ THz and $n_s\simeq 10^{12}-10^{13}$ cm$^{-2}$ as it was assumed in \cite{Khurgin14}). If, for example, $f=1$ THz and $n_s=10^{11}$ cm$^{-2}$, the typical nonlinear electric field (\ref{field}) is only 2.3 kV/cm which is three-four orders of magnitude smaller than in other materials. How large is then the nonlinear current, for example, the $n$-th harmonic current $j_{n\omega}$, if the fundamental harmonic field exceeds a few kV/cm? According to \cite{Mikhailov07e}, $j_{n\omega}\simeq 4en_sv_F/n\pi$ which gives (at $1$ THz and $n_s\simeq 10^{11}$ cm$^{-2}$) huge for a single atomic layer values of $j_{3\omega}\simeq 0.68$ A/cm, $j_{5\omega}\simeq 0.41$ A/cm, etc. At higher frequencies, e.g. at 30 THz as discussed in \cite{Khurgin14}, one does not need to restrict himself by the intra-band contribution only. Choosing the Fermi energy so that $\hbar\omega$ at 30 THz corresponds to $2E_F$ one gets $|\zeta_-(\omega)|\simeq 2000$. This is achieved at $E_F\simeq 60$ meV and $n_s\simeq 2.8\times 10^{11}$ cm$^{-2}$ which can be easily realized in graphene.

In summary, we have confirmed the statement of the original publication \cite{Mikhailov07e} that nonlinear graphene parameters are orders of magnitude larger than in many other nonlinear materials and that nonlinear properties of graphene can be used in and are tunable across a very broad frequency range. The physical reasons for this are the linear energy dispersion of graphene electrons, which works at low --- microwave, terahertz --- frequencies, and the presence of inter-band resonances relevant at high --- mid- and near-infrared, optical --- frequencies. Both these features  substantially reduce the electric fields needed for observation of nonlinear phenomena in graphene. We thus conclude, as opposed to Ref. \cite{Khurgin14}, that the nonlinear graphene optics and electrodynamics is a promising and encouraging field of research. 

The work has received funding from the European Union's Horizon 2020 research and innovation programme GrapheneCore1 under Grant Agreement No. 696656.

%\bibliography{../../../../../../../../../../../BIB-FILES/mikhailov,../../../../../../../../../../../BIB-FILES/thz,../../../../../../../../../../../BIB-FILES/graphene}

\end{document}